\documentclass[aps,prl,twocolumn,a4paper,superscriptaddress,citeautoscript]{revtex4-1}
\pdfoutput=1
\usepackage{graphicx}
\usepackage{amsmath, amsfonts}
\usepackage{amssymb}
\usepackage[colorlinks=true,linkcolor=blue,citecolor=blue]{hyperref}
\usepackage{color}

\begin{document}

\title{Temporal correlations and structural memory effects in break junction measurements}

\author{A.~Magyarkuti}
\affiliation{Department of Physics, Budapest University of Technology and Economics and \\ MTA-BME Condensed Matter Research Group, 1111 Budapest, Budafoki ut 8., Hungary}
\author{K.P.~Lauritzen}
\affiliation{Nano-Science Center and Department of Chemistry, University of Copenhagen, Universitetsparken 5, 2100 Copenhagen, Denmark}
\author{Z.~Balogh}
\affiliation{Department of Physics, Budapest University of Technology and Economics and \\ MTA-BME Condensed Matter Research Group, 1111 Budapest, Budafoki ut 8., Hungary}
\author{A.~Ny\'ary}
\affiliation{Department of Physics, Budapest University of Technology and Economics and \\ MTA-BME Condensed Matter Research Group, 1111 Budapest, Budafoki ut 8., Hungary}
\author{G.~M\'esz\'aros}
\affiliation{Research Centre for Natural Sciences, HAS, Magyar tudósok krt. 2, H-1117 Budapest, Hungary}
\author{P.~Makk}
\affiliation{Department of Physics, Budapest University of Technology and Economics and \\ MTA-BME Condensed Matter Research Group, 1111 Budapest, Budafoki ut 8., Hungary}
\author{G.C.~Solomon}
\affiliation{Nano-Science Center and Department of Chemistry, University of Copenhagen, Universitetsparken 5, 2100 Copenhagen, Denmark}
\author{A.~Halbritter}
\affiliation{Department of Physics, Budapest University of Technology and Economics and \\ MTA-BME Condensed Matter Research Group, 1111 Budapest, Budafoki ut 8., Hungary}

\date{\today}

\begin{abstract}
We review data analysis techniques that can be used to study temporal correlations among conductance traces in break junction measurements.
We show that temporal histograms are a simple but efficient tool to check the temporal homogeneity of the conductance traces, or to follow spontaneous or triggered temporal variations, like structural modifications in trained contacts, or the emergence of single-molecule signatures after molecule dosing. To statistically analyze the presence and the decay time of temporal correlations, we introduce shifted correlation plots. Finally, we demonstrate that correlations between opening and subsequent closing traces may indicate structural memory effects in atomic-sized metallic and molecular junctions.
Applying these methods on measured and simulated gold metallic contacts as a test system, we show that the surface diffusion induced flattening of the broken junctions helps to produce statistically independent conductance traces at room temperature, whereas at low temperature repeating tendencies are observed as long as the contacts are not closed to sufficiently high conductance setpoints. Applying opening-closing correlation analysis on Pt-CO-Pt single-molecule junctions, we demonstrate pronounced contact memory effects and recovery of the molecule for junctions breaking before atomic chains are formed. However, if chains are pulled the random relaxation of the chain and molecule after rupture prevents opening-closing correlations.
\end{abstract}

\maketitle
\section{Introduction}

Molecular electronics \cite{agrait03, molelectr, venkataraman_review} targets single-molecule devices with a much smaller active area than the resolution of present lithographic techniques. The construction of such small structures evidently relies on the atomic-scale self-organizing properties of the building blocks, and so it is important to investigate the reproducibility or the statistical variety of the emerging single-molecule structures. 

The break junction technique is an ideal tool for this statistical analysis: an atomic-sized metallic nanowire is repeatedly opened and closed (broken and reconnected), along which single-molecule bridges are formed again and again by the surrounding molecules \cite{agrait03, molelectr}.
The most basic break junction experiment is the measurement of the conductance along the opening and closing of metallic nanowires in molecular environment. Based on this repeated set of conductance vs. electrode separation measurements conductance histograms are constructed, which are regarded as a basic tool to determine the conductance of various single-molecule nanowires.

To go beyond the average conductance of a statistical ensemble of single-molecule nanowires, several additional techniques have also been introduced. 
A part of these techniques rely on the measurement of additional physical quantities \cite{venkataraman_review}, like noise \cite{brom99, Djukic06, Tal2016}, thermopower \cite{ludoph99a, thermopower1, thermopower2}, nonlinear conductance \cite{smit02, nonlin1, Scheer11}, force \cite{rubio96, force5, force6, force7}, etc., whereas another approach targets the advanced statistical analysis of the same conductance data from which the conductance histograms are constructed \cite{korrel1, ACS_NANO_CORREL, PtCO, AgCO, AuCO, yanson98, BPY2, 2DCDH1}.

In this paper we pose the question, to what extent are the conductance traces in break junction measurements statistically independent, and what kind of temporal correlations may be identified in the conductance data of breaking metallic and single-molecule nanowires? We demonstrate that temporal histograms serve as a useful tool to visualize temporal changes in the conductance traces of breaking nanowires. Next we introduce shifted correlation plots, a statistical tool to characterize temporal correlations. Finally we study correlations between opening and subsequent closing traces to investigate the \emph{structural memory} of atomic-sized metallic and single-molecule junctions. A part of the experimental data is compared with classical molecular dynamics simulations of breaking gold nanowires.    

\section{Results and Discussion}

\subsection{Conductance histograms and temporal conductance histograms}
\begin{figure*}[!htb]
\begin{center}
\includegraphics[width=1.3\columnwidth]{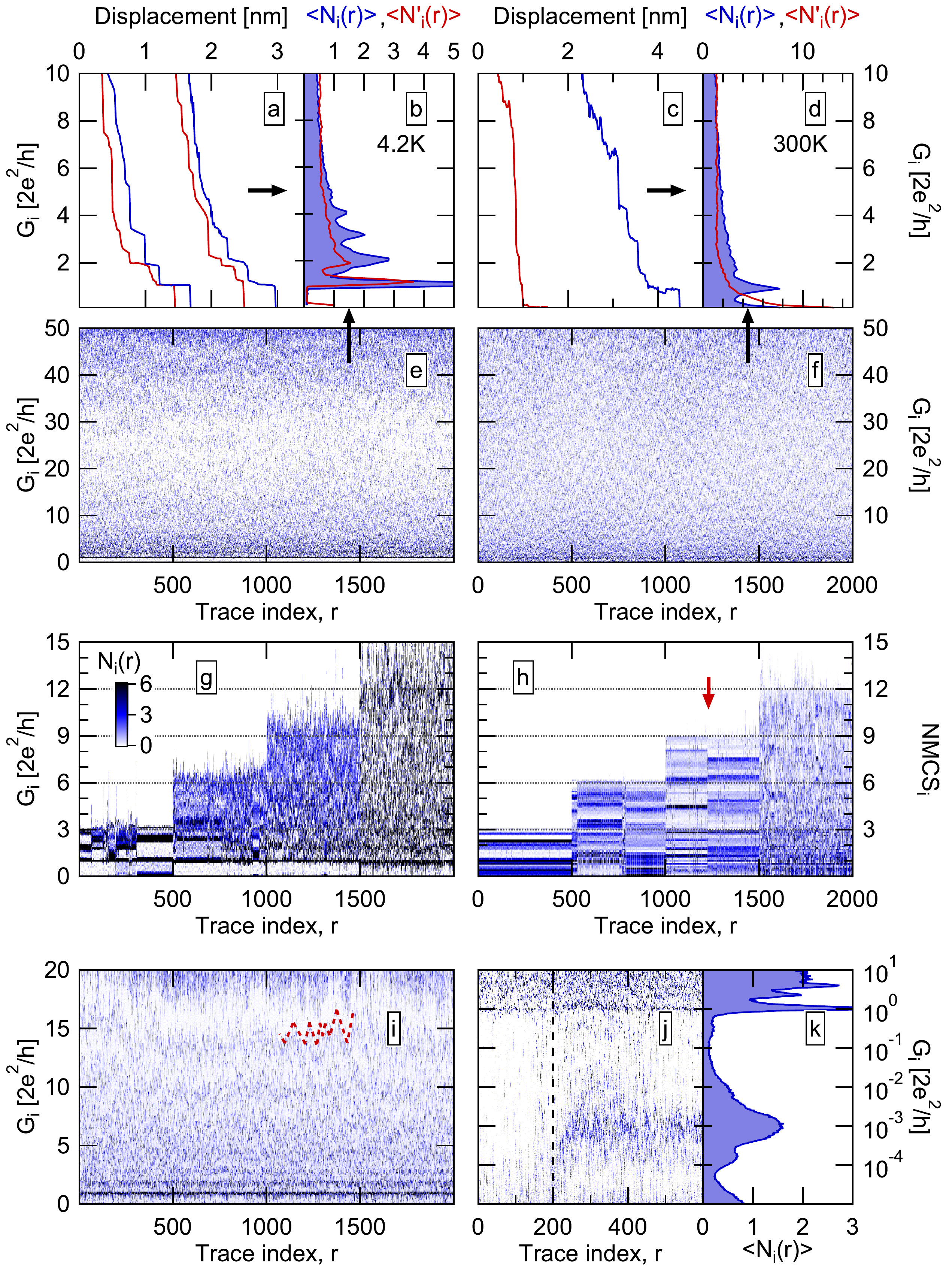}
\end{center}
\caption{\emph{Sample traces of breaking gold nanowires at $T=4.2\,$K (a) and room temperature (c), and the corresponding conductance histograms (b, d), and temporal histograms (e,f). The blue/red curves and histograms represent opening/closing traces. In both datasets, the junctions were closed to $50\,$G$_0$ setpoint. Temporal histograms of experimental (g) and simulated (h) conductance/normalized minimal cross section traces with closing setpoints of 3, 6, 9 and 12 G$_0$/NMCS, respectively. (i) Temporal histogram of gold opening traces recorded at $T=4.2\,$K with a closing setpoint of $20\,$G$_0$. The high conductance plateaus exhibit a temporal waving, as illustrated by the dashed line to guide the eye. (j) Temporal histogram of room temperature gold junctions demonstrating the in-situ dosing of 4,4'-bipyridine molecules. (k) Conductance histogram for the dataset on panel (j) after the molecular dosing. All temporal histograms use a similar colorscale as panel (g) with a scale fitted to the actual binning.}} \label{fig1}
\end{figure*}

Prior to introducing temporal correlation analysis techniques, we shortly summarize how conductance histograms are built.
Low temperature ($4.2\,$K) and room temperature sample traces of breaking gold nanowires are shown in Fig.~\ref{fig1}a and Fig.~\ref{fig1}c, respectively. The blue traces represent the variation of the conductance along the rupture of the junction, whereas the red traces demonstrate the conductance as the junction is closed again after the complete rupture. Based on these conductance data, single-trace conductance histograms are created, which are denoted by $N_i(r)$ and $N'_i(r)$ for the opening and closing trace respectively ($r$ is the trace index, $i$ is the bin index, and $N_i(r)$/$N'_i(r)$ denotes the number of data points in bin $i$ on the $r^\mathrm{th}$ opening/closing trace). The opening/closing conductance histogram of the entire dataset is obtained by averaging the single trace histograms for all traces ($H_i=\left<N_i(r)\right>_r$ and $H'_i=\left<N'_i(r)\right>_r$). The corresponding low and room temperature conductance histograms of the opening and closing traces respectively are plotted in  Fig.~\ref{fig1}b and Fig.~\ref{fig1}d. The conductance axes are scaled in the units of the conductance quantum, G$_0=2e^2/h$.

A reliable conductance histogram is based on a statistically diverse data set, such that any large enough subset of the traces (chosen without any specific selection criteria) would yield a histogram matching that of the entire dataset.
In many cases this is satisfied, but in Fig.~\ref{fig1} we also present some counterexamples using the most simple method of visualizing temporal imhomogeneities of conductance traces.

The quantity $N_i(r)$ for the entire dataset can be visualized as a two-dimensional image, resulting in a temporal histogram, as introduced in Ref.~\onlinecite{martinek16}. This is achieved by using conductance bins ($i$) along the vertical axis and the trace index ($r$) along the horizontal axis. The quantity $N_i(r)$ is then represented using a color scale as shown by the various temporal histograms in Fig.~\ref{fig1}. These color plots demonstrate the temporal evolution of the individual traces. To construct reliable conductance histograms a homogeneous temporal histogram is required, as shown in Fig.~\ref{fig1}e and Fig.~\ref{fig1}f for the low and room temperature opening traces of the datasets in Fig.~\ref{fig1}b and Fig.~\ref{fig1}d. In contrast, the temporal histograms in Figs.~\ref{fig1}g-j illustrate cases where this homogeneity requirement is not satisfied.

Fig.~\ref{fig1}g shows the low temperature measurement of Au opening traces where the junction is always closed to a predefined setpoint value, and then a hardware trigger stops the closing, and reopens the junction again. This means that each closing trace ends and each opening trace starts at a similar conductance value with minor overshoots. These temporal histograms exhibit segments with distinct horizontal lines  instead of the random scattering of the data indicating that a diverse set of junctions is not measured but rather measurements of a single junction structure is repeated multiple times. This tendency is typical in junctions closed to $3$ and $6\,$G$_0$ conductance (see first and second set of 1000 traces in Fig.~\ref{fig1}g), and is occasionally observed even at $9\,$G$_0$ setpoint, but at higher setpoints repeating traces are not likely. 

For comparison we have performed classical molecular dynamics simulations of breaking gold nanowires using Langevin dynamics with $T=4\,$K temperature, and calculated the minimal cross section data following the definition in Ref.~\onlinecite{Dreher2005}. The data are normalized according to the number of atoms in the smallest cross section (see methods). The temporal histogram of the normalized minimal cross section (NMCS) traces (Fig.~\ref{fig1}h) exhibits features similar to the experiments: up to a setpoint of $9\,$NMCS repeating traces are observed with occasional changes in the repeating junction evolution dynamics, whereas at a higher setpoint  a diverse set of traces is obtained. In gold junctions the NMCS is close to the conductance measured in the units of G$_0$\cite{agrait03}, so the threshold setpoint of observing repeating traces is similar for the experiment and the simulation. This agreement implies that the classical description with Langevin dynamics is sufficient to give us some insight into the repeating tendencies of breaking metallic nanowires. 

Based on the simulated traces, we have calculated the number of atoms that are changing their position significantly along an opening-closing cycle with a setpoint of $9\,$NMCS. An atom is regarded as a \emph{significantly displacing} atom, if its maximum displacement in the $x-y$ directions (perpendicular to the contact axis) along the whole opening-closing cycle is larger than $0.86\,$\AA , which is 30\% of the nearest neighbor distance in the Au fcc latice.
As an illustration we provide a video in the supporting information and also in Ref.~\onlinecite{Lauritzen2016_2} that shows the atomic rearrangements prior and after the sudden change denoted by the red arrow in Fig.~\ref{fig1}h, both from a front and from a back view. At the initial position with $9\,$NMCS, the two middle atomic layers consist of 6 and 8 atoms, respectively.  Naively, one would expect, that most of these atoms will show significant displacement, but the simulation yields only 4 significantly displacing atoms, which are color coded in the video. 3 out of these 4 atoms come from the atomic layer with 8 atoms. The first two opening and closing cycles correspond to precisely repeating NMCS traces, however the atomic rearrangements are not fully reproducible, rather some of the colored atoms are interchanged from the first to the second trace. At the third trace the atomic dynamics change significantly, reflected by a cut in the temporal histogram at the red arrow. The final two NMCS traces are again reproducible, but the colored atoms are again interchanged from trace to trace. A similar number of atoms is expected to account for the repeating traces in the experiments. 

Fig.~\ref{fig1}i shows an example where subsequent traces are not reproducible, rather a diverse dataset is observed. However, at higher conductances a waving of the typical plateau positions is observed on the scale of hundreds of traces, as indicated by the dashed line to guide the eye. This is attributed to the temporal variation of the typical junction structures, for example slow fluctuation of the dominant crystallographic orientation, opening angle of the junction, or other structural characteristics. A slow shift in the conductance values such as this, smears the characteristic peaks in the total histogram. 

Temporal histograms are also useful to follow the response of the conductance traces to some external change such as the dosing of molecules to the junction. Fig.~\ref{fig1}j shows an example in which the conductance traces of pure room temperature Au junctions are strongly changed when 4,4'-bipyridine molecules are introduced into the environment. The molecules are dosed in situ from a small quartz tube container, that is heated from outside by the tungsten spiral of a light bulb. After starting to heat the dosing tube (dahed line) the molecular signatures are already visible within 20 opening-closing cycles. (A part of this delay is attributed to the thermal inertia of the heating  tube, see methods for more details). The conductance histogram in Fig.~\ref{fig1}k demonstrates the result of molecule dosing: the characteristic double peak of 4,4'-bipyridine molecules is observed \cite{force6, BPY2}. This measurement demonstrates, that temporal histograms are also capable of following transient changes in the environment, with a temporal resolution down to the repetition rate of the traces.

\subsection{Shifted correlation plots}
To study the temporal correlations between different conductance traces one can generalize the correlation analysis method introduced in Refs.~\onlinecite{korrel1, ACS_NANO_CORREL}.  The correlation function between bin $i$ on a certain trace and bin $j$ on a trace shifted by $s$ breaking cycles is defined as
\small
\begin{equation}
\nonumber
C_{i,j}(s)=\frac{\left \langle \left(N_{i}(r)-\left< N_{i}(r)\right> \right) * \left(N_{j}(r+s)-\left< N_{j}(r+s)\right> \right) \right \rangle }{\sqrt{\left\langle \left(N_{i}(r)-\left< N_{i}(r)\right> \right)^{2} \right\rangle \left\langle \left(N_{j}(r+s)-\left< N_{j}(r+s)\right> \right)^{2} \right\rangle}}.
\end{equation}
\normalsize
The correlation function for bins on the same trace always shows distinct structures, including the evident perfect correlation at the diagonal ($C_{i,i}(s=0)=1$). In contrast the correlation function for shifted traces is expected to be zero for any ${i,j}$  pair, if the different traces are \emph{statistically independent}, i.e. any nonzero value reflects dependency between the traces, which may either come from reproducible segments, or slow structural variations, in accordance with Figs.~\ref{fig1}g,i. To visualize the temporal correlations we concentrate on the diagonal correlations, and plot $C_{i,i}(s)$ using the colorscale as a function of the shift $s$ and the conductance index $i$.  

\begin{figure}[!htb]
\begin{center}
\includegraphics[width=\columnwidth]{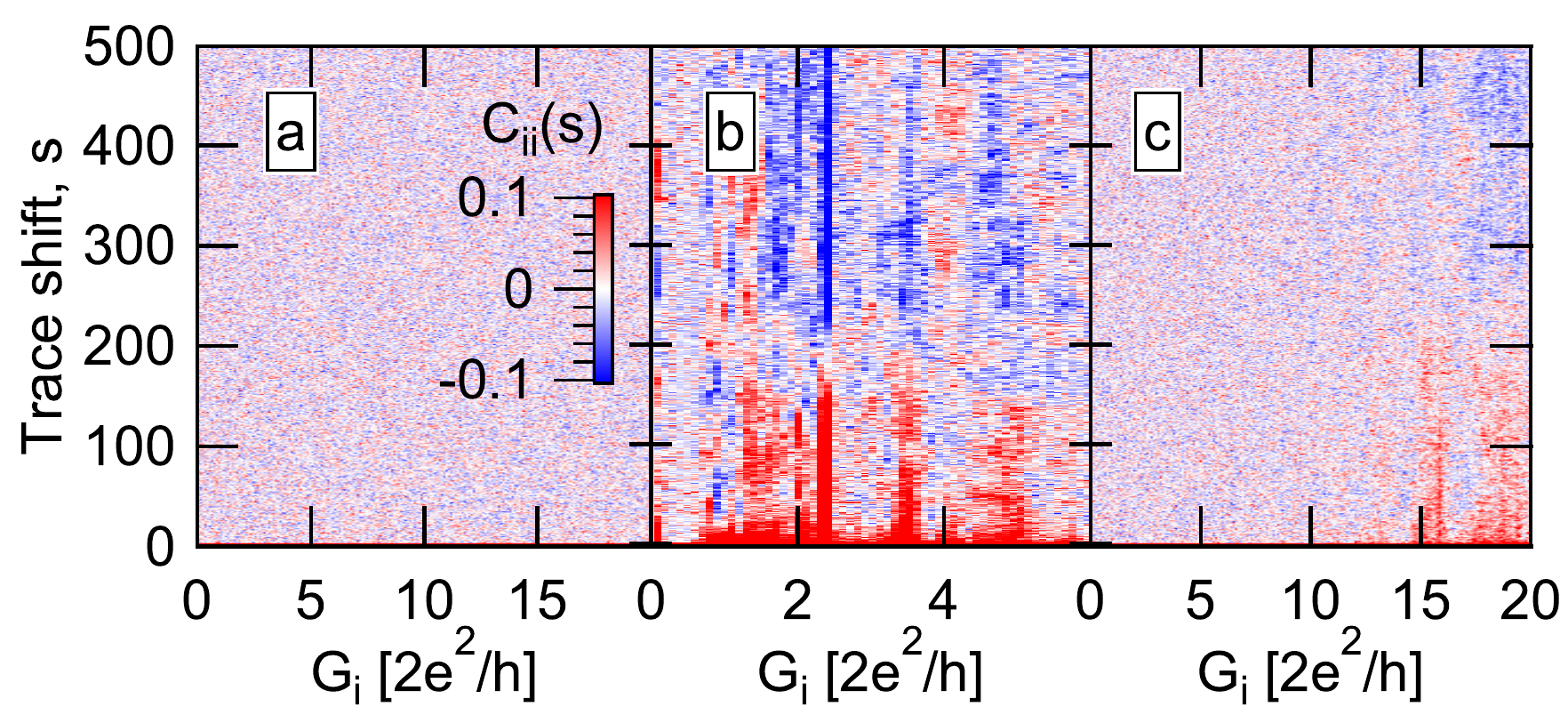}
\end{center}
\caption{\emph{The shifted diagonal correlation, $C_{i,i}(s)$ for the room temperature dataset in Fig.~\ref{fig1}e (a), for the repeating traces with $6\,$G$_0$ setpoint in Fig.~\ref{fig1}g (b), and for the traces with waving plateau positions in Fig.~\ref{fig1}i (c). In all panels the same colorscale is used.
}} \label{fig2}
\end{figure}

Fig.~\ref{fig2}a represents the shifted correlation plot for a statistically homogeneous room temperature Au data set, where the different traces are found to be independent of each other. It is to be noted, that at room temperature the junction surfaces flatten out after the complete rupture due to the strong surface diffusion of gold atoms \cite{surface_diff}. This is reflected by the large hysteresis between the opening and closing traces (see sample traces in Fig.~\ref{fig1}c). The hysteresis between jump out of contact along the opening trace and the jump to contact along the closing trace measured at $0.5\,$G$_0$ conductance has an average value of $2.9\,$nm. The jump to contact corresponds to a displacement where the prior opening trace exhibited $G=16.6\,$G$_0$ on average. This is in sharp contrast to low temperature measurements, were these values are respectively $1.09\,$nm and $2.85\,$G$_0$. 
Another consequence of surface diffusion is the absence of single atom peak in the closing histogram (red curve in Fig.~\ref{fig1}d): due to the flattened surfaces, the junction immediately jumps to higher conductance along the closing period. Due to this feature an experiment like the one in Fig.~\ref{fig1}g could not be conducted at room temperature: due to the flattening one cannot stop closing the junction at a few G$_0$ setpoint conductance. Therefore, at room temperature surface diffusion aids the statistical independence of the conductance traces.

In low temperature measurements, the independence of the traces is not obvious. At low closing setpoints even a few tens or hundreds of repeating traces introduce strong features in the shifted correlation plot, as demonstrated by Fig.~\ref{fig2}b. When repeating traces are excluded by higher closing setpoint, for example $20\,$G$_0$, a waving shift of the plateau positions is still observed (Fig.~\ref{fig1}i), which introduces significant correlations to the shifted correlation plot (Fig.~\ref{fig2}c) in the conductance regions where the waving is observed. The decay of these correlations with $s\approx200$ is consistent with the typical period of the plateau position waving. To eliminate such temporal correlations, one has to apply even higher closing setpoints, like $50-100\,$G$_0$, where the shifted correlation plot would not show any features, similar to Fig.~\ref{fig2}a.   
 
\subsection{Study of contact memory by cross-correlations between opening and closing traces}

The observation, that low temperature break junctions closed up to $9\,$G$_0$ conductance may exhibit completely reproducible opening traces (Fig.~\ref{fig1}g) tells us, that the complex motion of several atoms in the junction follows a highly reproducible tendency. To study this kind of \emph{contact memory} one can use a more general method, which does not require the controlled stoppage of the contact closing at a predefined value \cite{ACS_NANO_CORREL, AgCO}. Even, if the junction is closed to a much larger conductance, where subsequent opening traces would not show any correlation, the initial segment of the closing trace should show correlations with the previous opening trace. These correlations are expected to extend to conductances where a stopped closing followed by the reopening of the junction would yield repeating traces, or at least some repeating tendencies, like waving plateaus. On the other hand, the opening-closing correlation is somewhat suppressed due to the evident hysteresis between the opening and closing traces, therefore it is useful to study similarities between broader conductance intervals than a single conductance bin. Accordingly the opening-closing correlation function is defined as  
\small
\begin{equation}
\nonumber
C'_{i,j}=\frac{\left \langle \left(N_{i}(r)-\left<N_{i}(r)\right> \right) * \left(N'_{j}(r)-\left<N'_{j}(r)\right> \right) \right \rangle }{\sqrt{\left\langle \left(N_{i}(r)-\left<N_{i}(r)\right> \right)^{2} \right\rangle \left\langle \left(N'_{j}(r)-\left<N'_{j}(r)\right> \right)^{2} \right\rangle }},
\end{equation}
\normalsize
where $N$ and $N'$ respectively correspond to the opening and closing part of a trace pair \cite{AgCO}. If necessary the $N_i(r)$ values are replaced by a moving average along the conductance axis, $\bar{N}_i^a(r)=\left(\sum_{k=i-a..i+a}N_k(r)\right)/\left(2a+1\right)$ to enhance the resolution of possible correlations.

\begin{figure}[!htb]
\begin{center}
\includegraphics[width=\columnwidth]{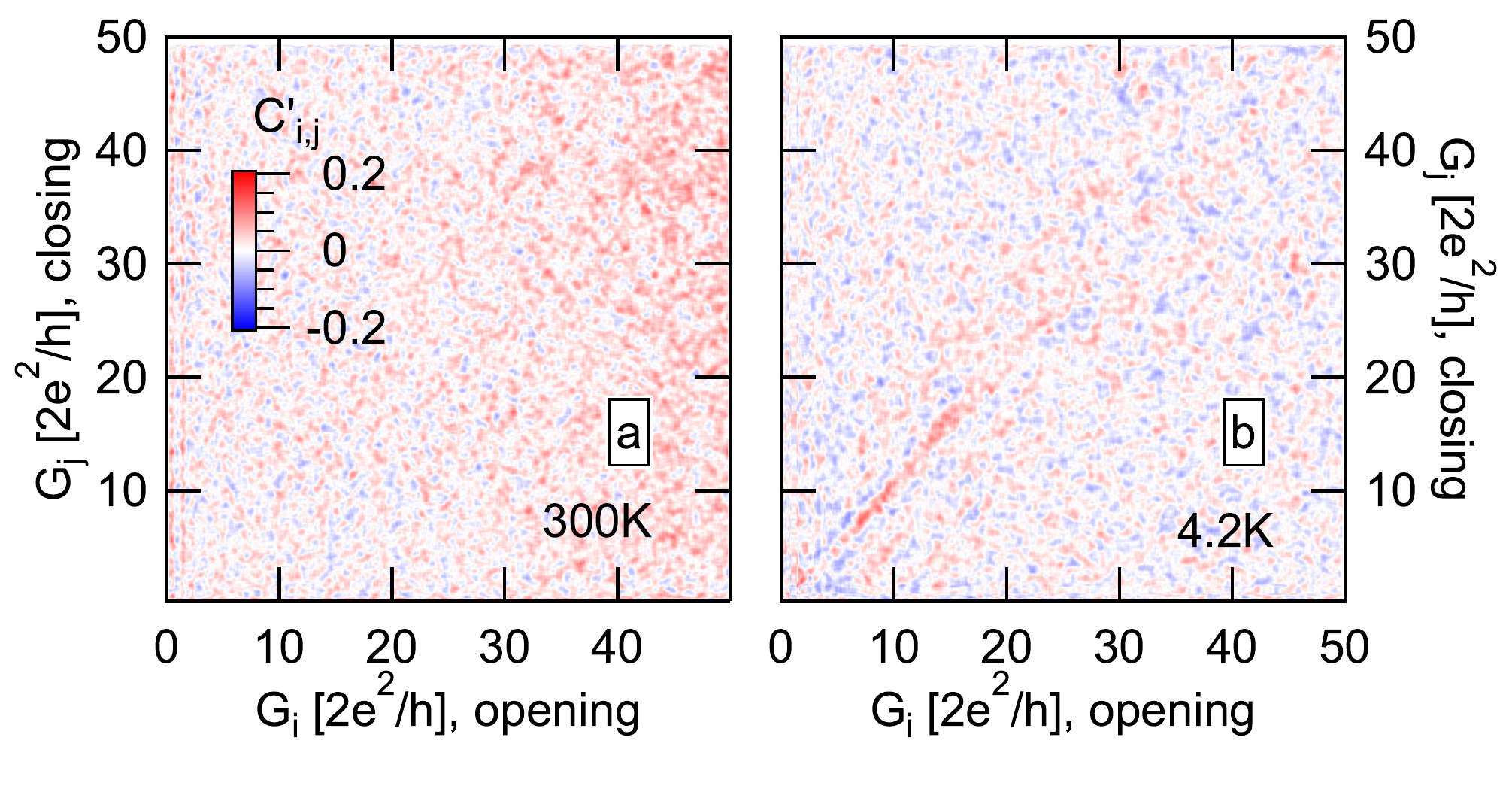}
\end{center}
\caption{\emph{(a/b) Opening-closing correlation plots, $C'_{i,j}$ for the room temperature/$T=4.2\,$K homogeneous datasets in Fig.~\ref{fig1}f/e, respectively.  A bin size of $0.1\,$G$_0$, a moving average of $5$ bins ($a=2$), and the same colorscale is used in both panels. 
}} \label{fig3}
\end{figure}

In room temperature measurements, pronounced opening-closing correlations are not common due to the surface diffusion induced flattening of the contact surfaces, as demonstrated by Fig.~\ref{fig3}a. 

At cryogenic temperatures the apexes of the broken junction are more rigid, as reflected by the small hysteresis between the jump out of contact and jump to contact points (Fig.~\ref{fig1}a), the formation of single-atom contacts in the closing histogram (Fig.~\ref{fig1}b), and the tendency for repeating traces (Fig.~\ref{fig1}g).  The opening-closing correlation plot for the statistically homogeneous dataset in Fig.~\ref{fig1}e indeed shows positive correlations at the diagonal (Fig.~\ref{fig3}b). Up to the correlated opening-closing conductances some kind of contact memory is preserved, which would be reflected by repeating tendencies on a subsequent opening trace, if the contact was not closed further than these correlated opening-closing conductance values. 


\begin{figure}[!htb]
\begin{center}
\includegraphics[width=\columnwidth]{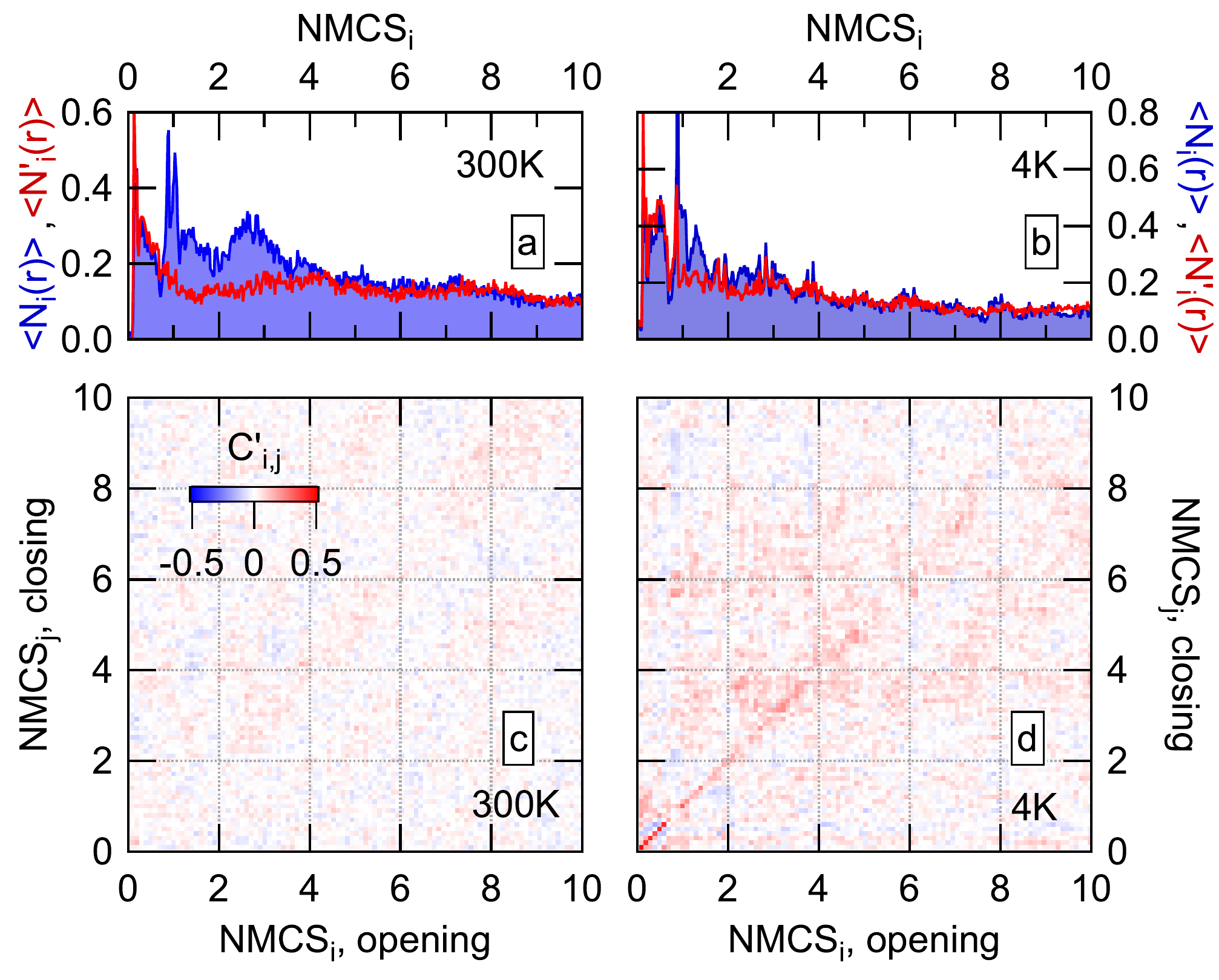}
\end{center}
\caption{\emph{(a,b): Opening (blue) and closing (red) conductance histograms of simulated NMCS traces with with $T=300\,$K and $T=4\,$K tempearature, respectively. (c,d): The opening-closig correlation plots, $C'_{i,j}$ for the same simulated datasets. Both panels use the same colorscale, a bin size of $0.1$, and no averaging ($a=0$).
}} \label{fig4}
\end{figure}

Similar features are also observed in the traces simulated by Langevin dynamics: the room temperature simulations reproduce the surface diffusion induced contact flattening, yielding the absence of the $1\,$NMCS peak in the closing histogram, and the absence of any features in the opening-closing correlation plot (Fig.~\ref{fig4}a,b), whereas the $T=4\,$K simulations reproduce the clear $1\,$G$_0$ peak in the closing histogram, and the opening-closing correlations around the diagonal  (Fig.~\ref{fig4}c,d).

It is noted, that distinct features of shifted correlation plots, $C_{i,j}(s)$ (like the result of waving plateaus) are usually inherited by the opening-closing correlation plots as well. Therefore, the opening-closing correlations are the most interesting in datasets, where the shifted correlations are absent, as it is the case in Figs.~\ref{fig3} and \ref{fig4}. 

\subsection{Opening-closing correlations in single molecule junctions}

Finally we investigate the evolution of single-molecule junctions by opening-closing correlation analysis. If the molecule is likely to stay upright in the junction after the rupture and so a similar molecular configuration is formed as the electrodes are closed, a positively correlated region is expected around the diagonal of the opening-closing correlation plot. The absence of correlations points to the random rearrangement of the junction after the rupture. In our previous work we have already studied such features for CO molecules contacted by Ag electrodes \cite{AgCO}. Now we demonstrate another single-molecule system where both the recovery of the single-molecule junction and the loss of correlations due to random atomic rearrangements can be observed. To this end, we analyze Pt-CO-Pt single molecule junctions, for which the evolution of the contact during rupture was described in detail based on conditional two-dimensional conductance-displacement histograms \cite{PtCO}. These junctions exhibit two distinct single molecule configurations ($G\approx 1.1\,$G$_0$ and $G\approx 0.5\,$G$_0$) as the single-atom junction ($G\approx 2.1\,$G$_0$) is stretched further in CO environment. The higher/lower conductance single-molecule configuration corresponds to a CO molecule perpendicular/parallel to the contact axis. From the plateau length histograms, it was shown that not only pure Pt junctions exhibit monoatomic chain formation \cite{smit01}, but both the perpendicular and parallel CO molecules bind strongly enough to the junction to pull monoatomic Pt chains out of the electrodes \cite{PtCO}. It is found that the perpendicular molecule infiltrates the junction prior to monoatomic chain formation, and then the chains are pulled through the molecule, such that the CO either stays in the perpendicular orientation or changes to parallel as the chain is pulled.

\begin{figure}[!htb]
\begin{center}
\includegraphics[width=\columnwidth]{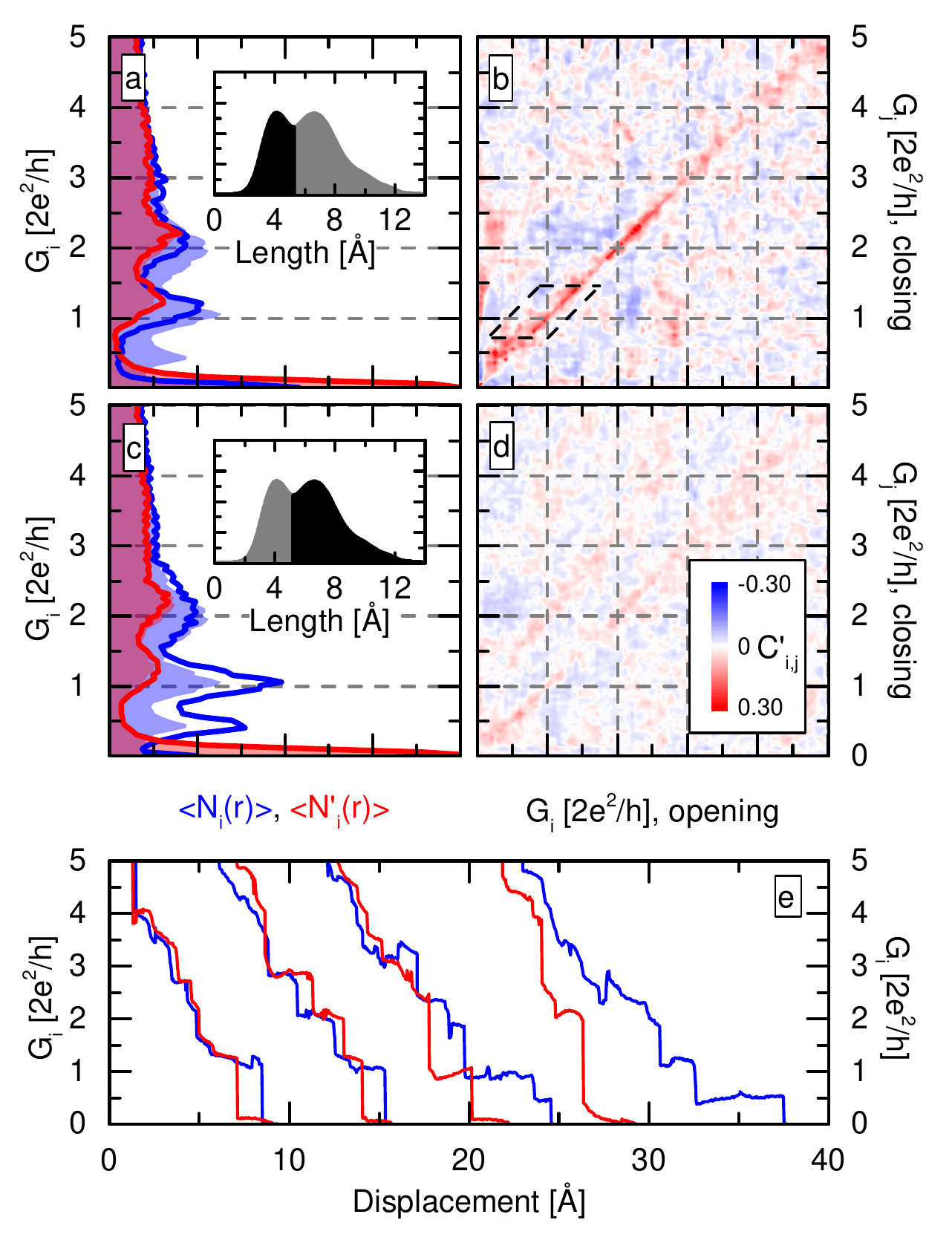}
\end{center}
\caption{\emph{(a-d) Conductance histograms and opening-closing correlation plots for Pt-CO-Pt single molecule junctions for selected traces, where the opening part does not show atomic chain formation (a,b) or at least one Pt chain atom is pulled out from the electrodes (c,d). The black regions of the plateaus length histograms in the insets define the length regions of these selections. The opening/closing histograms of the selected traces are shown by blue/red lines, whereas the semitransparent blue/red area graphs show the opening/closing histograms of the entire dataset as a reference. The two correlation plots use the same colorscale shown in panel (d). Panel (e) shows sample traces from the datasets of panel (a,b)/(c,d) (see first and second trace/third and fourth trace). The first and second trace illustrate that the closing trace may follow similar plateaus as the opening one, which is not exclusive, but frequently observed on traces without chain formation. If chains are pulled the absence of reproducible tendencies is common (third and fourth sample trace).
}} \label{fig5}
\end{figure}

In Fig.~\ref{fig5} we study the correlations between the opening and closing traces separately for traces exhibiting/not exhibiting atomic chain formation along the opening period (see panels (c,d)/(a,b)). To concentrate on single-molecule junctions, we only analyze the traces breaking from any molecular configuration, but not from a single-atom contact ($74\%$ of all traces, see Ref.~\onlinecite{PtCO} for details). To separate the traces we use the plateau length histogram constructed for the conductance region spanning the single atom peak and both molecular peaks ($0.2-2.75\,$G$_0$) along the opening part. The insets show this plateau length histogram, with the length region of the corresponding selection denoted by black. Panels (a) and (c) show the opening/closing conductance histograms for the such selected traces by blue/red lines. As a reference the opening/closing histograms for the entire dataset are shown by semitransparent blue/red area graphs.

The opening-closing correlation plot for the traces without chain formation (Fig.~\ref{fig5}b) exhibits pronounced positive correlations around the diagonal, indicating a strong memory effect. This indicates that the closing trace is likely to reproduce the opening one both around the molecular conductance and at higher atomic conductances. This is indeed observed in a significant portion of the traces, as shown in the first and second opening-closing sample trace in panel (e). It is also seen that in the conductance region of perpendicular CO junctions, the positively correlated region (see the encircled area) is somewhat offset from the diagonal. This is related to the smaller conductance of a stretched molecular junction on the opening trace than the relaxed one along the closing trace,  which is also exhibited by the offset of the corresponding opening/closing histogram peaks. Note, that the traces without chain formation do not exhibit the low conductance parallel CO junction \cite{PtCO}.

In contrast, traces, where atomic chains are pulled exhibit imperceptible correlations (Fig.~\ref{fig5}d). This is related to the random relaxation of the chain atoms and the CO molecule to the electrodes after rupture, which induces a strong stochasticity to the traces, such that the memory effects are lost, i.e. no reproducibility is found between the opening-closing trace pair (see the third and fourth sample trace in panel (g)). 

It is to be noted, that the closing conductance histogram is practically the same for the entire dataset (semitransparent red area graph), for the chain pulling traces (red line in Fig.~\ref{fig5}c) and for the traces where chains are not pulled (red line in Fig.~\ref{fig5}c): regardless of these selections the single molecule junction with perpendicular CO is formed with similar probability and weight in all cases. However, if chains are not pulled, the correlation analysis and the sample traces show that the closing trace regularly yields the \emph{same} molecular configuration as the opening trace, i.e. no significant rearrangement occurs when the electrodes are disconnected. 

\section{Conclusions}
Temporal inhomogeneities in conductance traces may be regarded as a disturbing factor, raising questions about the reliability of conductance histograms, but these can also serve as a useful source of information. As a simplest example, even a small portion of repeating traces may introduce unrepresentative peaks to the conductance histogram, but one can even take advantage of the fully reproducible motion of several atoms, e.g. if atomic-scale memories are investigated  \cite{Geresdi11, Scheer13, BPY2}. We have demonstrated that the simple method of temporal histograms is an efficient way of visualizing temporal variations in break junction data. This can be used to check the temporal homogeneity of the traces, but it can also be used to follow spontaneous or triggered temporal variations, like the waving of conductance plateaus, the appearance of repeating traces as the contact is trained, or the emergence of molecular plateaus after molecular dosing. With this method one may even follow the temporal dynamics of in-situ chemical reactions as some external parameters are tuned. We have also shown that the temporal homogeneity of the conductance traces can be checked by shifted correlation plots, where any feature shows that the traces are not statistically independent of each other, and the correlation time is indicated by the decay of the features as the function of the shift number, $s$.
We have also shown that correlations between opening and subsequent closing traces may indicate structural memory effects. In metallic contacts, this can be used to study when the structural characteristics of the breaking junction are preserved along the closing period, whereas in molecular junctions it can highlight when the same molecule can be recovered after rupture.

Using our specific test system of gold metallic contacts, we have demonstrated that in room temperature measurements the surface diffusion induced flattening of the broken junctions helps to produce statistically independent conductance traces, whereas at low temperatures the rigid contacts are likely to show repeating traces and waving plateaus as long as the closing setpoint is not high enough. The former features were also demonstrated by classical molecular dynamics simulations. The use of opening-closing correlation analysis was also demonstrated on Pt-CO-Pt single molecule junctions. Pronounced contact memory effects were found if the single-molecule junction breaks before atomic chains are formed. However, if chains are pulled through the CO molecule, random relaxation of the chain atoms and the CO molecule after rupture prevents opening-closing correlations.

\section{Acknowledgements}
The authors are thankful to L. Venkataraman and J. Martinek for useful discussions. A. H. acknowledges the financial support from the National Research, Development and Innovation Office (K105735 and K119797 research grants). K.P.L. and G.C.S acknowledge financial support from the Danish Council for Independent Research, Natural Sciences and the Carlsberg Foundation.

\section*{Methods}
\subsection*{Experimental methods}
The measurements were performed with self-designed mechanically controllable break junction setups either at room or at liquid helium temperature. High purity Au or Pt wires ($0.1$mm diameter) were fixed on the top of a bending beam. Between the two fixing points, a notch was created reducing the diameter of the junctions to $10$-$30\,\mu$m. The fine tuning of the electrode separation was performed by a piezo actuator with a reduction of $\approx1:50-1:100$ due to the MCBJ geometry. The conductance versus electrode separation traces were measured by repeatedly opening and closing the junction. Starting from the initial position the junction was opened by $\approx20\,$nm displacement with $20-100\,$nm/s speed, and afterwards it was closed with the same speed till the setpoint conductance. Applying a bias voltage of $100\,$mV the conductance was measured through a FEMTO DLPCA-200 linear current amplifier or by a self-built logarithmic current amplifier \cite{logIV} in the case of molecular measurements in Fig.~\ref{fig1}j. 

In the measurement shown by Fig.~\ref{fig1}j the molecules were dosed from a quartz tube by heating the tube from outside with the tungsten spiral of a light bulb. 
Due to the thermal inertia of the dosing system in the room temperature measurements around $10$ seconds is required to heat up the quartz container such that a significant amount of molecules are evaporated. Meanwhile a conductance trace is recorded in every second. If the tube heating was stopped after $9$ seconds, no molecular signatures are observed on the traces. However an extra second of heating is immediately reflected by a molecular plateau on the next trace, and 10 traces later already almost all curves show molecular signature. After the $10\,s$ dosing period the heating is switched off, but there are already enough molecules close to the junction, so that the molecular plateaus are observed for many thousands of conductance traces. 

In the low temperature measurements on Pt-CO-Pt single-molecule junctions the CO molecules were dosed through a stainless steel tube going from a room temperature vacuum flange to the
vicinity of the cryogenic temperature sample \cite{PtCO}.

In the low temperature measurements the electrode displacements were calibrated based on the period of peaks in the plateaus length histograms of pure Au or Pt atomic chains \cite{untiedt02,smit01}. At room temperature the exponential decay of tunneling curves was fitted to that in STM break junction measurements with gold, where the direct displacement of the piezo actuator is known. 

\subsection*{Theoretical simulations}
An initial gold junction was built from an fcc crystal with dimensions: 8 atoms by 8 atoms by 12 atoms.
Dividing the junction into bottom lead (3 layers), wire (6 layers), and top lead (3 layers) along the $z$ axis, the wire was trimmed to a cylinder with radius 7 \AA{}, discarding atoms outside this cylinder (The initial structure is provided in the supporting information and in ref. \onlinecite{Lauritzen2016}).

The repeated opening and closing of this junction was simulated using the Atomic Simulation Enviroment (ASE)\cite{Bahn2002}. Effective medium theory (EMT) \cite{Jacobsen1996} was used as the energy calculator for the Langevin dynamics (with timestep=5.0 fs and friction=0.1 inverse atomic units of time and the temperature set at either 4 K or 300 K).

During the dynamics, only the atoms in the wire were allowed to move. Every 100 MD steps the top lead was displaced 1 pm (stretching the junction). After the total displacement of the top lead equalled 2 nm, the direction of the displacement was changed to contract the junction again.
The contraction continued until the minimal cross-sectional area (MCS) reached a threshold value (indicating 3, 6, 9 or 12 atoms in the MCS), at which point the direction of displacement changed again. This cycle was repeated 500 times.

For the results presented in Fig. \ref{fig4}, the threshold for when to stop contracting was computed from the distance between the top and bottom leads instead.
When this distance was equal to the distance between the leads in the initial structure,  the simulation was stopped. The simulation was repeated 1000 times, every time starting from the same initial structure.

The MCS was calculated following the definition in \cite{Dreher2005}. Briefly, the junction is cut into slices, the volume of the atoms in each slice is computed and the area of a cylinder with the same volume (and same height as the slices) is then defined as the minimal cross-sectional area.
The normalization factor (to transform MCS area to NMCS) was found by taking 1/3th of the MCS area on a structure with 3 atoms as the minimal cross-section.

\bibliographystyle{prsty}
\bibliography{manuscript_stochasticity_arxiv}

\end{document}